\documentclass[a4paper,11pt]{article}
\usepackage{pos}
\usepackage{amsmath}
\usepackage{mathtools}
\usepackage{subfigure}
\usepackage{cleveref}

\definecolor{mymagenta}{RGB}{200, 0, 100}
\definecolor{myblue}{RGB}{45, 48, 146}

\graphicspath{{plots/}}

\title{Defining Canonical Momenta for Discretised SU$(2)$ Gauge Fields}

\author[a]{Marco Garofalo}
\author[b]{Tobias Hartung}
\author[c]{Karl Jansen}
\author[d]{Johann Ostmeyer}
\author[a]{Simone Romiti}
\author*[a]{Carsten Urbach}

\affiliation[a]{Helmholtz-Institut für Strahlen- und Kernphysik \&
  Bethe Center for Theoretical Physics,\\ Rheinische Friedrich-Wilhelms-Universität Bonn, Nussallee 14-16, 53115 Bonn, Germany}
\affiliation[b]{Northeastern University - London, Devon House, St Katharine Docks, London, E1W 1LP, United Kingdom}
\affiliation[c]{NIC, DESY Zeuthen, Platanenallee 6, 15738 Zeuthen, Germany}
\affiliation[d]{Department of Mathematical Sciences,
	University of Liverpool, United Kingdom}

\emailAdd{urbach@hiskp.uni-bonn.de}

\abstract{In this proceeding contribution we discuss how to define
  canonical momenta for SU$(N)$ lattice gauge theories in the
  Hamiltonian formalism in a basis where the gauge field operators are
  diagonal. For an explicit discretisation of SU$(2)$ we construct the
  momenta and check the violation of the fundamental commutation
  relations.
}

\newcommand{\ri}{\mathrm{i}}

\FullConference{%
The 39th International Symposium on Lattice Field Theory,\\
8th-13th August, 2022,\\
Rheinische Friedrich-Wilhelms-Universität Bonn, Bonn, Germany
}

\begin{document}

\maketitle

\hypertarget{introduction}{%
\section{Introduction}\label{introduction}}

The Hamiltonian of lattice gauge theories was formulated by Kogut and
Susskind in Ref.~\cite{Kogut:1974ag} already in 1974. For an SU$(N)$
lattice gauge theory it reads in generic form
\begin{equation}
  \label{eq:hamiltonian}
  \hat H =\ \frac{g_0^2}{4}\sum_{\mathbf{x},c,k} \left(\hat{L}_{c,k}^2(\mathbf{x}) +
  \hat{R}_{c,k}^2(\mathbf{x})\right) + \frac{1}{2g_0^2}\sum_{\mathbf{x},k<l} \mathrm{Tr}\,
  \mathrm{Re}\, \hat{P}_{kl}(\mathbf{x})\,,
\end{equation}
with $g_0$ the (bare) gauge coupling constant and the lattice
spacing set to $a=1$.
Note that without discretised gauge fields $\hat L^2$ and $\hat R^2$
are identical. 
Here, $\mathbf{x}$ are the coordinates in a $n$-dimensional, spatial
lattice with lattice spacing $a$
\[
\Lambda\ =\ \{\mathbf{x}: x_k = 0, a, 2a, \ldots, (L-1)a\}\,,
\]
with periodic boundary conditions and $k=1,2,\ldots,n$. $c$ labels the
colour index of the group 
SU$(N)$. The trace is taken in colour space and $\hat{P}_{kl}(x)$ are
plaquette operators
\[
\hat{P}_{kl}(\mathbf{x})\ =\ \hat{U}_k(\mathbf{x})\, \hat{U}_l(\mathbf{x}+\hat k)\,
\hat{U}^\dagger_k(\mathbf{x}+\hat l)\, \hat{U}^\dagger_l(\mathbf{x})\,,
\]
with $\hat k$ a vector of length $a$ in direction $k$. The elements
$\hat u_{ij}:\mathcal{H}\to\mathcal{H}$ of
$\hat{U}_k\in\mathrm{SU}(N)$ represent the gauge field operators in
direction $k$ and the $\hat{L}_{x,k}$ and $\hat{R}_{c,k}$ are the corresponding
canonical momenta. In the following the spatial coordinates
$\mathbf{x}$ and the directions will not be relevant and, thus, we
will drop them. 

Given the generators $t_c$ of the group SU$(N)$, the elements of the
$\hat{U}$ and their canonical momenta are defined via the commutation
relation 
\begin{equation}
  \label{eq:commUL}
        [\hat{L}_c, \hat{U}_{mn}]\ =\ (t_c)_{mj}\, \hat{U}_{jn}\,,
        \qquad [\hat{R}_c, \hat{U}_{mn}]\ =\ \hat{U}_{mj}(t_c)_{jn}\,.
\end{equation}
Moreover, the $\hat{L}_c$ resemble the group structure
\begin{equation}
  \label{eq:commLL}
        [\hat{L}_a, \hat{L}_b] = f_{abc} \hat{L}_c\,,
\end{equation}
with the the structure constants $f_{abc}$ of the corresponding Lie
algebra, and likewise the $\hat{R}_c$.

If this formalism is to be implemented using tensor networks or on
future digital quantum computers, a discretisation scheme is needed
with a corresponding truncation scheme for the Hilbert space. And the
race is on to find the most efficient way to implement this
discretisation. For different schemes on the market see for instance
Ref.~\cite{Davoudi:2020yln}.

Most of the existing discretisations of the Hamiltonian have in common
that they work in a basis where the kinetic / electric part of $\hat
H$ is diagonal. The magnetic part is then obtained for instance by a
character expansion. In this proceeding we explore the possibility to
work in a basis where the gauge field operators are diagonal: this
might be advantageous in particular regions of parameter space, see
Refs.~\cite{Paulson:2020zjd,Haase:2020kaj} for a discussion in Abelian
U$(1)$ theory.

\hypertarget{state-space}{%
\section{State Space}\label{state-space}}

To simplify the discussion and to be concrete, we resort to the
special case of SU$(2)$ in the following with generators given by the
Pauli matrices and colour indices $c=1,2,3$. We will chose states
$|U\rangle\in\mathcal{H}$ in a Hilbert space $\mathcal{H}$ which are
eigenstates of the operators $\hat{U}$ in the following sense:
parametrise an SU$(2)$ matrix using three real valued parameters $y_0, 
y_1, y_2$ with
\[
\begin{pmatrix}
  y_0 + \ri y_1 & y_2 + \ri y_3 \\
  -y_2 + \ri y_3 & y_0 - \ri y_1\\
\end{pmatrix}\in \mathrm{SU}(2)\,,\qquad y_3^2 = 1- \sum_{i=0}^2 y_i^2\,.
\]
Now define operators $\hat y_j:\mathcal{H}\to\mathcal{H}$ by the following action
\[
\hat y_j |U\rangle\ =\ y_j |U\rangle\,.
\]
Defining also 
\[
\begin{split}
  \hat u_{00} = \hat y_0 + \ri \hat y_1\,,\quad \hat u_{01} = \hat y_2
  + \ri \hat y_3\,,\\
  \hat u_{10} = -\hat y_2 + \ri \hat y_3\,,\quad \hat u_{11} = \hat y_0
  - \ri \hat y_1\,,\\
\end{split}
\]
we can set for $\hat U:\mathcal{H}\to\mathcal{H}$
\[
\hat U =
\begin{pmatrix}
  \hat u_{00} & \hat u_{01}\\
  \hat u_{10} & \hat u_{11}\\
\end{pmatrix}\,.
\]
Therefore, the $y_{1,2,3}$ can be regarded as quantum numbers
labelling the states $|U\rangle$ which are simultaneous eigenstates of
operators $\hat y_{1,2,3}$.

Formally, we can define the momenta as Lie derivatives:
\begin{equation}
  \label{eq:momenta}
  \hat{L}_c\, f(\hat{U})\ = -\ri \frac{\mathrm{d}}{\mathrm{d}\alpha}\,
  f\left(e^{\ri\,\alpha t_c}\, \hat{U}\right)|_{\alpha = 0}\,,\qquad
  \hat{R}_c\, f(\hat{U})\ = -\ri \frac{\mathrm{d}}{\mathrm{d}\alpha}\,
  f\left(\hat{U}\, e^{\ri\,\alpha t_c}\right)|_{\alpha = 0}
\end{equation}
for a function $f(\hat{U})$.

We make the Hilbert space finite by using one of the partitionings we
proposed in Ref.~\cite{Hartung:2022hoz}. These partitionings define a
finite set of group elements $G_\mathrm{M} = \{y(\mu): \mu=1, \ldots,
N(M)\}$,  which are asymptotically isotropic and dense in
SU$(2)$ depending on a parameter $M\in\mathbb{N}$. The continuum
group is approached with $M\to\infty$. The mean distance roughly goes
like $1/M$ and the number of elements $N(M)\sim M^3$.

\hypertarget{discretising-the-derivative-in-su2}{%
\section{\texorpdfstring{Discretising the derivative in
SU\((2)\)}{Discretising the derivative in SU(2)}}\label{discretising-the-derivative-in-su2}}

With one of the aforementioned partitionings the discretisation of the
operators $\hat U$ and the state space is straightforward. However,
the discretisation of the canonical momenta is more involved. The
finite difference operation
\begin{equation}
\begin{split}
  \frac{1}{\alpha}\left(f(e^{\ri\alpha t_a}\, \hat U) - f(\hat U)\right)\ &=\ 
  \frac{1}{\alpha}\left(f(\hat U) + \alpha \hat L_a f(\hat U) -
  f(\hat U)
  +O(\alpha^2)\right)\\
  &=\ \hat L_a f(\hat U)\, + O(\alpha)
\end{split}
\end{equation}
in direction $a$ is a natural way to implement the
discretisation. Thus, we need to reconstruct the directional
derivative from the existing neighbouring elements in $G_\mathrm{M}$. Let $U\in
G_\mathrm{M}$ be one element for which we desire to define $\hat L_a$.
Let us chose a specific representation and $f(\hat U) = \hat U$. 
Then, one
can find three neighbours $V_i\,,\ i=1,2,3$ of this element
$U$. Moreover, there are three $W_i\in \mathrm{SU}(2)$
\begin{equation}
  \label{eq:Vi}
  V_i(U) = W_i\, U\qquad \Leftrightarrow\qquad W_i =  V_i U^{-1} =
  \exp(\ri \alpha_b^i t_b)
\end{equation}
with $\alpha_b^i\in\mathbb{R}$, $b=1,2,3$ and $i=1,2,3$.
Now, with additional real parameters $\gamma_i\in\mathbb{R}$ we can 
expand as follows
\[
\begin{split}
\gamma_1 W_1 U + \gamma_2 W_2 U + \gamma_3 W_3 U &\approx
\left(\gamma_1(1+\mathrm{i}\alpha_b^1 t_b) + \gamma_2(1+\mathrm{i}\alpha_b^2 t_b )+
\gamma_3(1+\mathrm{i}\alpha_b^3 t_b)\right) U\\
&= \sum_i \gamma_i U +\ri\sum_{i,b} \gamma_i\alpha_b^i t_b U\,.
\end{split}
\]
Therefore, one needs to determine the parameters
$\gamma_i$ such that
\begin{equation}
  \sum_{i,b} \gamma_i\alpha_b^i t_b = t_a\,,
\end{equation}
because then
\[
\begin{split}
  -\left(\sum_{i=1}^n \gamma_i\right) U & + \gamma_1 V_1 + \gamma_2 V_2 + \gamma_3 V_3  \\
  &= -\left(\sum \gamma_i\right) U+ \gamma_1 W_1 U + \gamma_2 W_2 U +
  \gamma_3 W_3 U \\
  &\approx \mathrm{i}\left(\gamma_1\alpha_b^1
  t_b + \gamma_2\alpha_b^2 t_b +
  \gamma_3\alpha_b^3 t_b\right) U\\
  &= \mathrm{i} t_a U\,.
\end{split}
\]
Thus, the algorithm reads
\begin{enumerate}
\item find three next neighbours $V_i\in G_\mathrm{M}$ of one element $U\in G_\mathrm{M}$, then
  compute the $W_i$ as defined above and the three vectors
  $\alpha_b\in\mathbb{R}^3$. 
\item combine the column vectors $\alpha_b$ into a $3\times3$ matrix and solve
  \[
  e_a = \gamma \cdot (\alpha_1 \alpha_2 \alpha_3)
  \]
  for vector $\gamma\in\mathbb{R}^3$ with $e_a\in\mathbb{R}^3$ the unit vector in
  direction $a$. 
  
\item the only non-zero elements of the discrete operator $\hat
  L_a\in\mathbb{R}^{N(M)\times N(M)}$ are then given by
  \[
  (\hat L_a)_{\# U,\, \# V_i(U)} = \gamma_i\,,\qquad (\hat L_a)_{\#
    U,\,\# U} = -\sum_i \gamma_i
  \]
  with $\# V_i(U)$ the index of $V_i(U)$ and $\# U$ the index of $U$
  in $G_\mathrm{M}$. 
\end{enumerate}
For determining also $R_a$ the algorithm only needs to be modified by
replacing \cref{eq:Vi} by
\[
V_i = U W_i\,,\quad \Leftrightarrow\quad W_i = U^{-1} V_i\,.
\]

\hypertarget{test-of-the-commutation-relation-with-u}{%
\section{\texorpdfstring{Test of the Commutation Relations}{Test of the Commutation Relations}}\label{test-of-the-commutation-relation-with-u}}

\begin{figure}
  \centering
  \includegraphics[width=.9\textwidth]{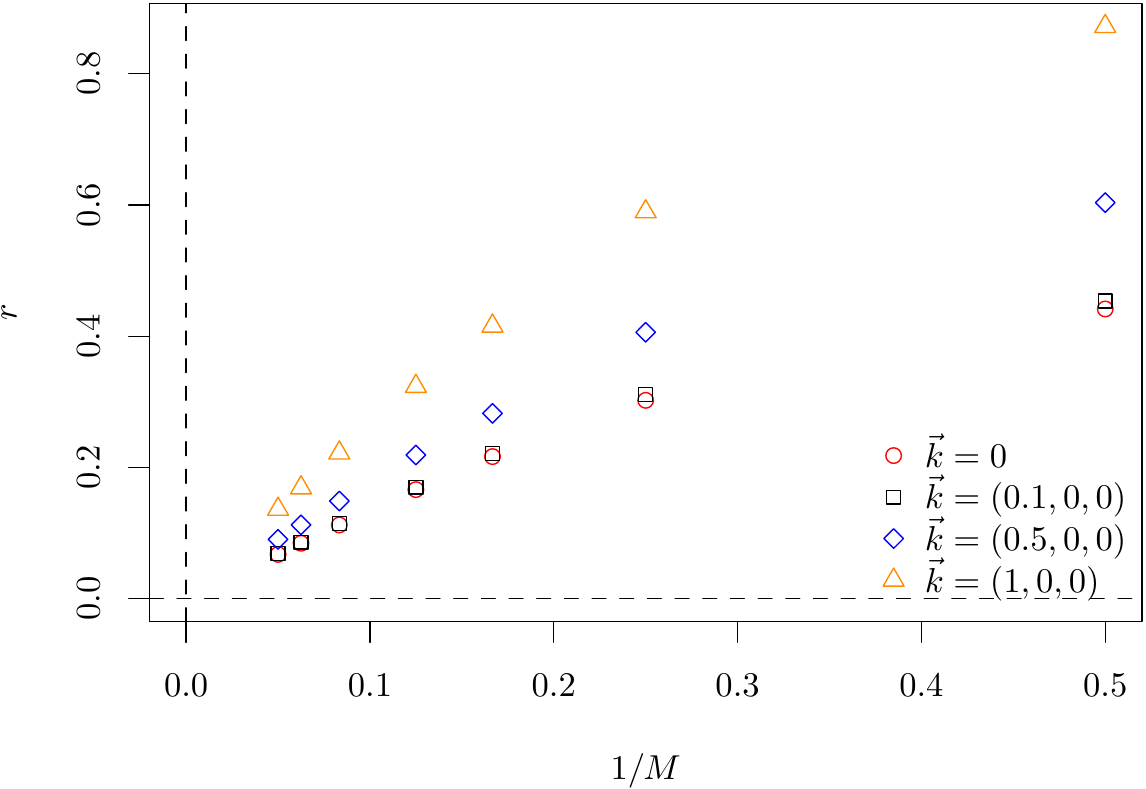}
  \caption{We plot $r$ for the commutator \cref{eq:zLU} as defined in
    the text as a function of $1/M$ 
    for different Fourier vectors $\vec k$.}
  \label{fig:commLU}
\end{figure}

For the test of this discretisation we check whether the commutation
relations \cref{eq:commUL,eq:commLL} are approximately fulfilled by
the discretised operators defined above. For this purpose we chose
what we call linear partitioning in Ref.~\cite{Hartung:2022hoz}, which
is defined by the following set of points
\begin{equation}
  \begin{split}
    G_\mathrm{M} &\coloneqq \left\{\frac{1}{J}\left(s_0j_0,\dots,s_3j_3\right)\right.\\
    &\left|\,\sum_{i=0}^3 j_i=M,\;\forall i\in \{0,\dots,3\}:\,s_i\in\{\pm1\},\,j_i\in\mathbb{N}\right\}\,,
  \end{split}\label{eq:linear_discretisation}
\end{equation}
with
\[
J \coloneqq \sqrt{\sum_{i=0}^3 j_i^2}\,.
\]
This is directly related to the aforementioned parametrization of
SU$(2)$ via $(s_0j_0, s_1j_1,s_2j_2,s_3j_3)/J \equiv (y_0, y_1, y_2, y_3)$.

With the above definitions of $\hat L$ and $\hat R$ it is ensured that
if applied to a constant vector one obtains zero. Much like in the one
dimensional case of a finite difference operator, we expect $\hat L$
and $\hat R$ to work best if applied to slowly varying vectors in the
algebra. This is why we define the equivalent of Fourier modes in the
algebra denoted by $v(\vec k)$. 
Since the convergence is correct to \(O(\alpha)\) for each element of
$G_\mathrm{M}$ separately, we compute
\begin{equation}
  \label{eq:zLU}
  z\ =\ \left([L_a, U_{jl}] -(t_a)_{ji} U_{il}\right)\cdot v(\vec k) 
\end{equation}
and then the mean deviation as
\begin{equation}
  r\ =\ \frac{1}{N(M)}\sum |z_i|
\end{equation}
with \(N(M)\) the number of points in the set $G_\mathrm{M}$. Note
that one could equivalently use
\[
r^\prime = \frac{\langle v| z \rangle}{\langle v|v\rangle}\,.
\]
In \cref{fig:commLU} we show the result of our test for different
Fourier vectors $\vec k$ by plotting $r$ as a function of $1/M$.
One can observe that $|r|$ increases at fixed $M$ with the modulus of
$\vec k$. Moreover, for all vectors $\vec k$ we see convergence of
$r\to0$ with $M\to\infty$. We also note that the average deviation is
not particularly small for the $M$-values considered here.

\begin{figure}
  \centering
  \includegraphics[width=.9\textwidth]{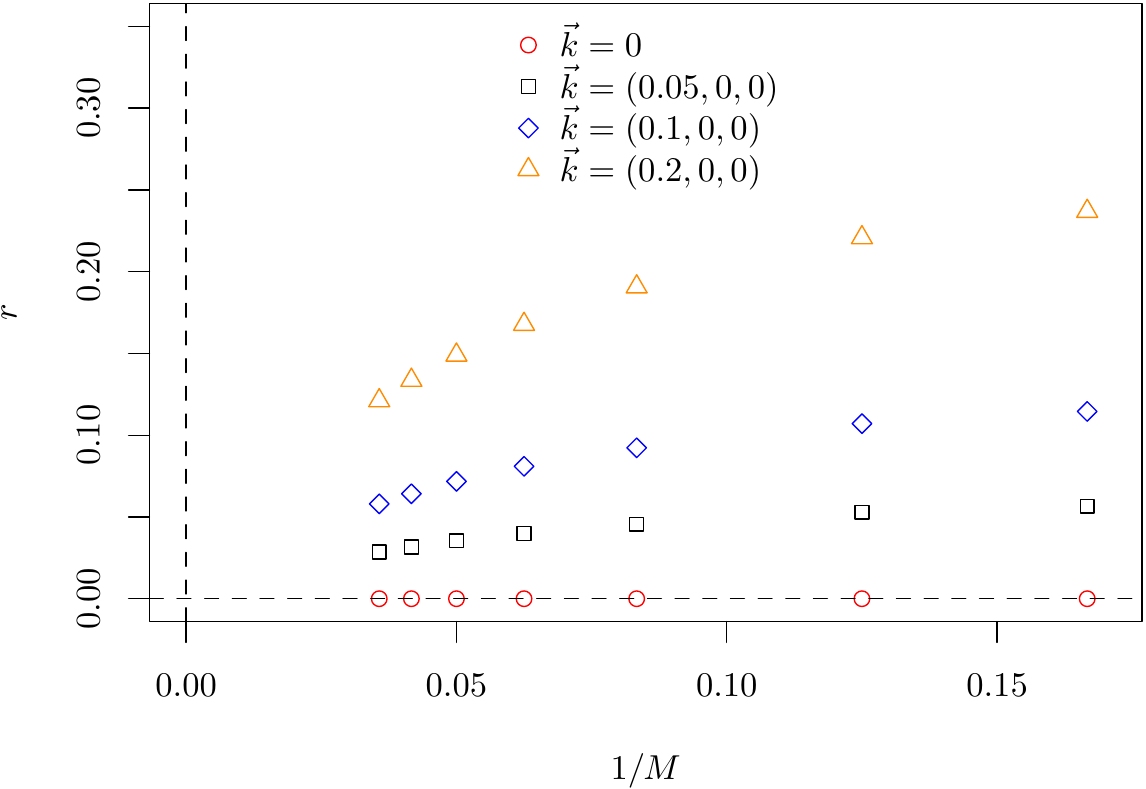}
  \caption{We plot $r$ for the commutator \cref{eq:zLL} as defined in
    the text as a function of $1/M$ 
    for different Fourier vectors $\vec k$.}
  \label{fig:commLL}
\end{figure}

In \cref{fig:commLL} we again show $r$ as a function of $1/M$, but
this time we define
\begin{equation}
  \label{eq:zLL}
  z\ =\ \left([L_a, L_b]\ +\ 2 \mathrm{i}\,\epsilon_{abc}\,L_c\right)\cdot v(\vec k) 
\end{equation}
with the appropriate $f_{abc}$ for SU$(2)$. Note that the scale of the
$x$-axis is different compared to \cref{fig:commLU} and also the
Fourier vectors are different with smaller $|\vec k|$-values than in
\cref{fig:commLU}. 

First of all, for $\vec k =0$, $r$ vanishes independently of $M$. This
is due to the fact that $L_a v(0) = 0$ per construction.
We observe that for this
commutator the convergence appears to be slower: only at $1/M < 0.2$
convergence towards zero becomes plausible, even though probably at
least a factor two larger values of $M$ are needed to reliably
establish this observation. 

\hypertarget{summary-and-outlook}{%
\section{\texorpdfstring{Summary and Outlook}{Summary and Outlook}}\label{summary-and-outlook}}

In this proceeding contribution we have discussed how to define
canonical momenta for discretised SU$(N)$ gauge fields. We have tested
our discretisation scheme with one particular gauge group
discretisation and we observe that in the limit of continuous group
the exact commutation relations are recovered. The particular construction
discussed here for SU$(2)$ can be generalised to SU$(3)$.

The obvious next steps are the investigation of the spectrum of the
free theory using the discretised momenta and to compare to other
discretisation schemes. And, of course, an implementation of the
Hamiltonian for a digital quantum computer must be explored.

\begin{acknowledgments}
  We thank A.~Crippa, G.~Clemente and J.~Haase for helpful discussions.
  This work is supported by the Deutsche
Forschungsgemeinschaft (DFG, German Research Foundation) and the NSFC through the funds provided to the Sino-German
Collaborative Research Center CRC 110 “Symmetries
and the Emergence of Structure in QCD” (DFG Project-ID 196253076 -
TRR 110, NSFC Grant No.~12070131001)
  as well as the STFC Consolidated Grant ST/T000988/1.
  The open source software package R~\cite{R:2019} has been used.
\end{acknowledgments}

\bibliographystyle{h-physrev5}
\bibliography{bibliography}

\end{document}